\documentclass[12pt]{appolb}
\usepackage{epsfig}

\newcommand{\be}{\begin{eqnarray}}
\newcommand{\ee}{\end{eqnarray}}
\newcommand{\nee}{\nonumber\end{eqnarray}}
\newcommand{\nn}{\nonumber}

\def\figa
{
\begin{figure}[h]\vspace{-0.5cm}
\hspace{-1cm}
\epsfig{figure=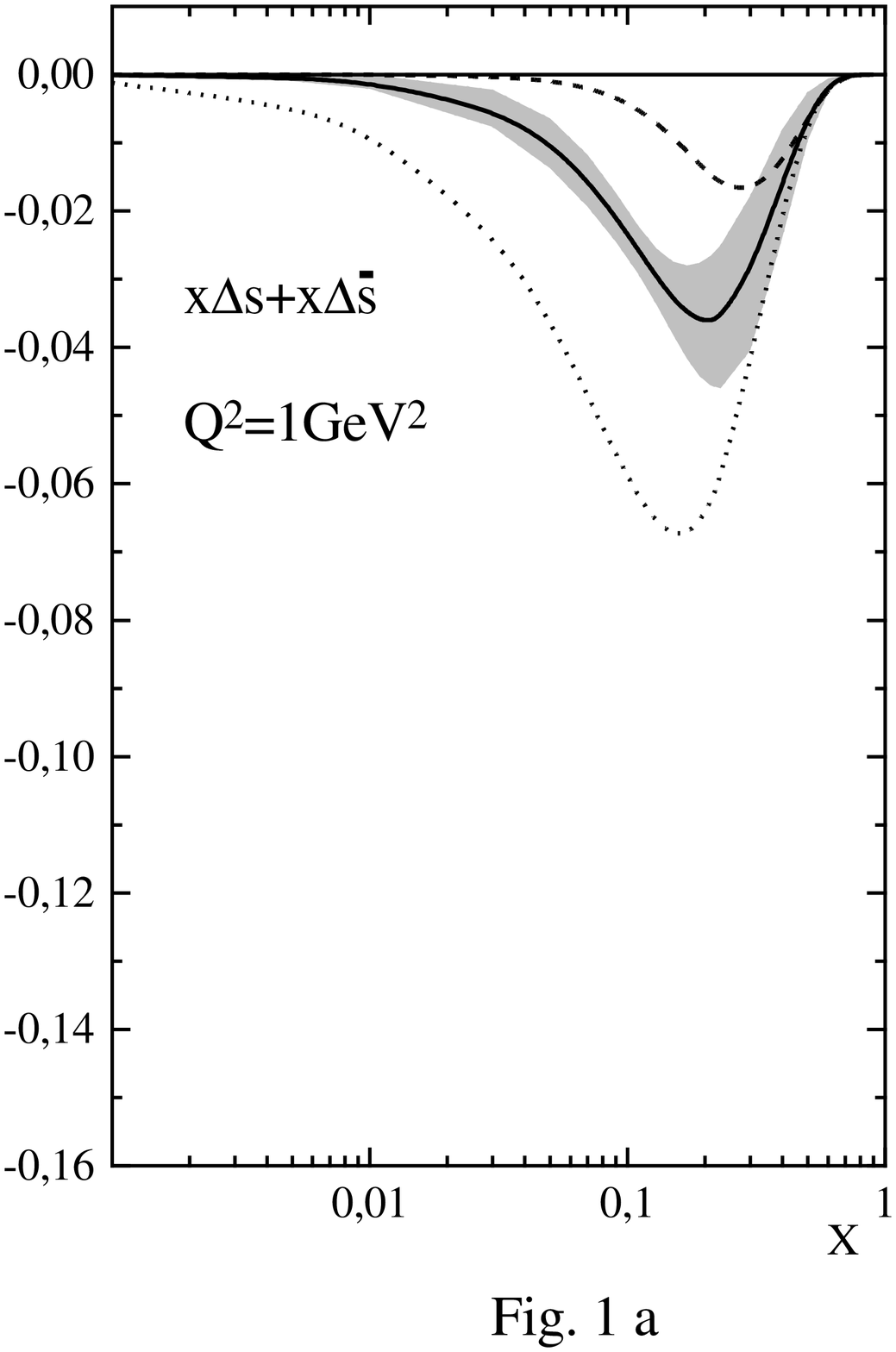,width=8.5cm}
\epsfig{figure=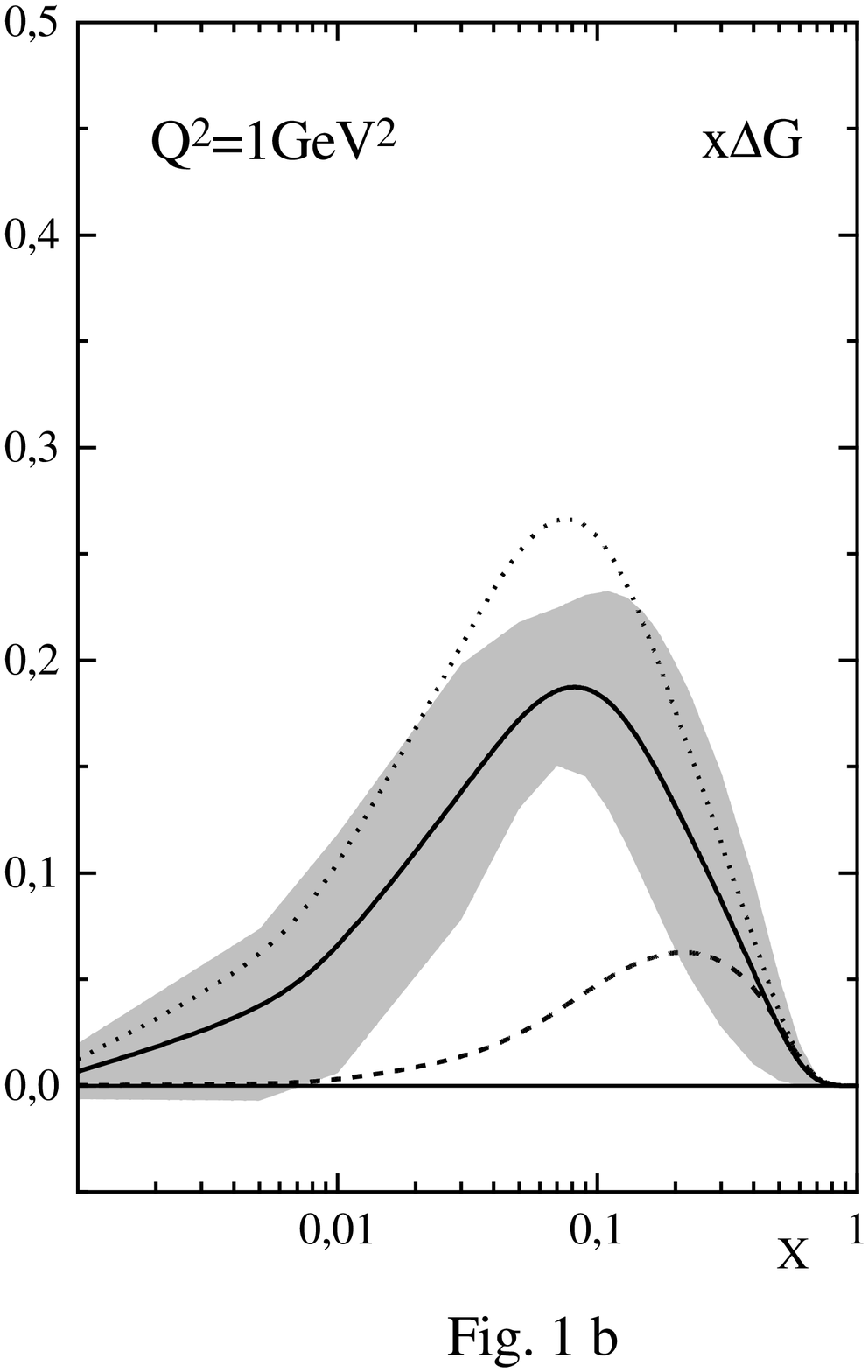,width=8.5cm}
\vspace{-0.5cm}
\caption{
Variation in $\Delta s + \Delta\bar s$ and  $\Delta G$ arising from
varying the value of $a_8$ in (\ref{a8}) away from its SU(3) value of 0.58
(solid line and error band); $a_8=0.86$ (dotted
line);  $a_8=0.40$ (dashed line).
}
\end{figure}
}

\def\figb
{
\begin{figure}[h]\vspace{-0.5cm}
\hspace{-1cm}
\epsfig{figure=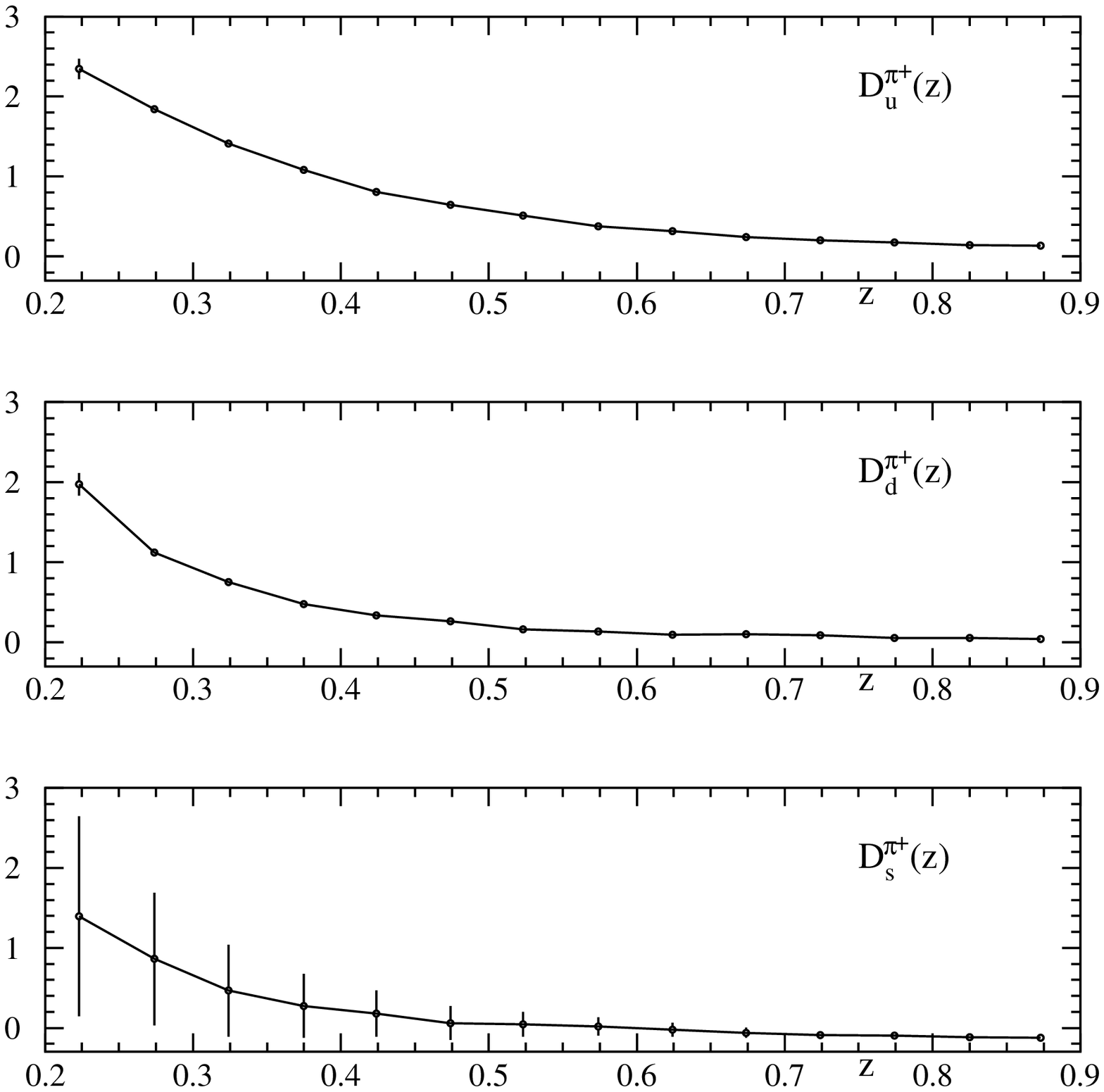,width=8.5cm}
\epsfig{figure=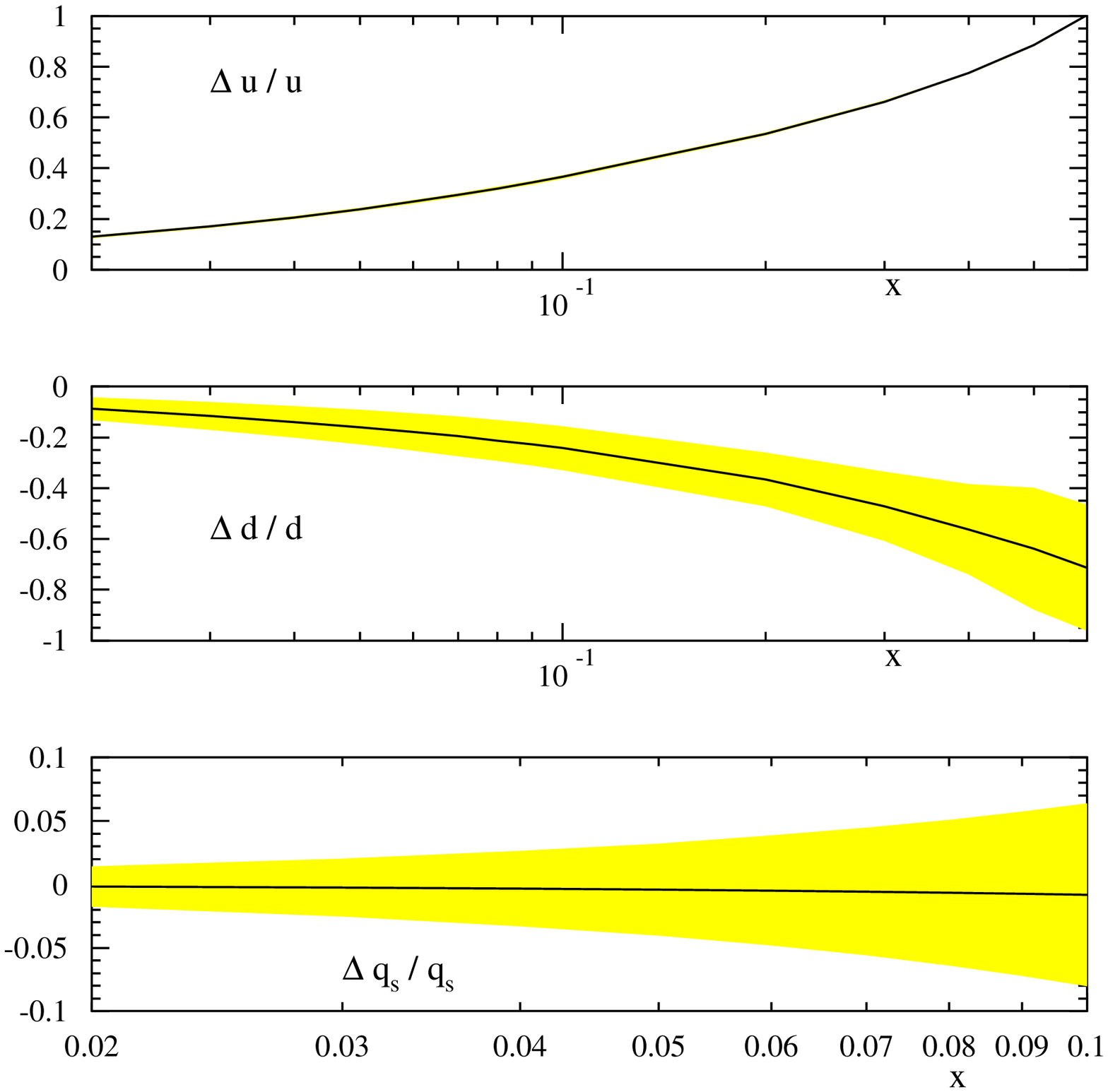,width=8.5cm}
\vspace{-0.5cm}
\caption{
The extracted fragmentation functions (left) with their
 uncertainties.
Polarized parton densities (right)
extracted from fake, error-free
 data, as explained in the text - the uncertainties arise solely
 from the realistic experimental errors on our  FFs.
}
\end{figure}
}

\begin{document}

\setcounter{page}{1}
\begin{center}
{\large\bf  Fragmentation functions and their role in determining the polarized parton densities}

\vspace{1cm} S. Kretzer$^{1}$, E. Leader$^{2}$, E.
Christova$^{3\dag}$,
 \vskip 5mm
 {\small (1) {\it Michigan State University, East Lansing, MI}
\\
(2) {\it Vrije Universiteit, Amsterdam\\
Permanent address: Imperial College, London }
\\
(3) {\it Institute of Nuclear Research
and Nuclear Energy, Sofia }
 \\
 $\dag$ {\it E-mail:
kretzer@pa.msu.edu, e.leader@ic.ac.uk, echristo@inrne.bas.bg  }}
\end{center}

\begin{center}
Presented by E. Leader
\end{center}

\begin{center}
We extract the pion fragmentation functions and their
uncertainties from a judicious choice of $e^+e^-$ and
semi-inclusive DIS data. These are used to study the error
propagation in the extraction of polarized parton densities from
semi-inclusive DIS asymmetries. We conclude that the uncertainties on polarized
PDs have been underestimated in the past.
\end{center}



\section{Introduction \label{sect:intro}}

 Flavour separation of the unpolarized parton
densities (PDs) relies heavily on the neutrino CC reactions.
Determination of the gluon density relies on the large range of
$Q^2$ available. Neither of these possibilities exist for the
polarized PDs. Thus, although the data from polarized DIS studies
are now of superb quality, they, in principle, only provide
information on the combinations $\Delta q(x) + \Delta \bar q(x)$
and on $\Delta G(x)$, and both $\Delta s(x) + \Delta \bar s(x)$
and  $\Delta G(x)$ are poorly known.

Ultimately, a neutrino factory and TESLA-N would remedy this, but
for the next decade progress will rely upon HERMES and COMPASS
semi-inclusive (SIDIS) data.

Extraction of the polarized PDs from polarized SIDIS requires a
good knowledge of fragmentation functions (FFs). We derive values
of the pion FFs and a realistic assessment of their uncertainties.
We then study how the uncertainties of the FFs generate errors on
the extracted polarized PDs. We conclude that previous studies
have underestimated these errors.

For simplicity  we employ a LO formalism~\cite{we1}. A strategy
for a simplified NLO treatment is given in \cite{we2}.


\section{ Polarized DIS -- current status\label{sect:DIS}}

Remember that, in principle, we can only obtain information on
$\Delta u(x) + \Delta \bar u(x)$, $\Delta d(x) + \Delta \bar
d(x)$, $\Delta s(x) + \Delta \bar s(x)$ and  $\Delta G(x)$ from
present day polarized DIS. Moreover, whereas $(\Delta u + \Delta
\bar u)-(\Delta d + \Delta \bar d)$ is directly determined from
$g_1^p(x,Q^2)-g_1^n(x,Q^2)$, the separation of $\Delta G$, $\Delta
s + \Delta \bar s$ and $(\Delta u + \Delta \bar u)+(\Delta d +
\Delta \bar d)$ from each other relies partly on evolution -- not
very efficacious with the presently available range of $Q^2$ --
and partly on the hyperon $\beta$-decay sum rule \be a_8 \equiv
\int_0^1  dx \left\{ \Delta u + \Delta \bar u+\Delta d + \Delta
\bar d -2\left( \Delta s + \Delta \bar s\right)  \right\} =
3F-D=0.58\cdot\label{a8} \ee All the modern analyses~\cite{DIS}
agree quite well on the determination of $(\Delta u + \Delta \bar
u)$ and $(\Delta d + \Delta \bar d)$ (results are often presented
for $\Delta u_V$, $\Delta d_V$ and $\Delta_{sea}$, but this is
quite artificial and usually incorporates some assumptions like
$\Delta \bar u= \Delta\bar d =\Delta s$ etc.), but there is
substantial disagreement about $(\Delta s + \Delta\bar s)$ and
$\Delta G$; though all suggest a non-zero, positive first moment
$\int dx \Delta G(x)$.

Eq. (\ref{a8}) assumes that the $\beta$-decays of
the hyperon octet respect flavour SU(3) symmetry, a belief
 challenged by some authors. We have therefore studied the
effect of varying the value of $a_8$, (\ref{a8}), between the
extreme estimates 0.40 and 0.86 found in literature~\cite{a8}.
Very likely this is too extreme to be realistic, but the huge
effect on the poorly determined $(\Delta s + \Delta \bar s)$ and
$\Delta G$, as seen in Fig.1
\figa
is instructive. It underlines the
importance of other sources of information on the polarized PDs and
this, in the foreseeable future, means polarized SIDIS at HERMES
and COMPASS, which, as mentioned,  requires an accurate knowledge
of the FFs to which we now turn.

\section{Extraction of the Fragmentation Functions}

For a long time it was thought that the majority of FFs were well
determined by the $e^+e^-\to hX$ reaction. Then in 2000/2001 three
papers appeared which all described essentially the same $e^+e^-$
data equally well, but with significantly different individual
flavour FFs~\cite{FFs}. In other words (and in retrospect this is
not surprising) $e^+e^-\to hX$ does not provide a sensitive method
of flavour separation.

Despite this we have been able to extract the pion FFs with
considerable accuracy, by combining recent HERMES data on
$\pi^\pm$ multiplicities on protons with a judicious choice of
information from $e^+e^-\to \pi^\pm X$~\cite{we3}. The latter
involves a subtle trick which allows us to avoid the above
mentioned ambiguities in the FFs derived from $e^+e^-\to hX$.

There are 3 independent FFs for pions: $D_u^{\pi^+}$,
$D_d^{\pi^+}$ and $D_s^{\pi^+}$. All the others can be related to
these via charge conjugation invariance or isotopic spin
invariance. The HERMES data on $\pi^\pm$ multiplicities on protons
provides us with 2 equations for 3 unknowns (the $\tilde\sigma$
differ from the cross sections $\sigma$ by a standard kinematic
factor -- see \cite{we3}):
 \be
 \tilde\sigma^h =\frac{4}{9}\left[
u(x)D_u^h(z) + \bar u(x)D_{\bar u}^h(z)\right] + \frac{1}{9}\left[
d(x)D_d^h(z) + \bar d(x)D_{\bar d}^h(z) + s(x)D_s^h(z) + \bar
s(x)D_{\bar s}^h(z)\right]\label{sigmatilde}
 \ee
  where $h=\pi^\pm$
and where we assume known unpolarized PDs.

To obtain a third relation we note that in $e^+e^-\to h X$ at
low $Q^2$ the coupling to $q\bar q$ pairs is electromagnetic, with
\be
e_u^2 =\frac{4}{9}\,>\,e_d^2 \,=\,e_s^2 \,=\,\frac{1}{9}\cdot\label{eu}
\ee
However at $Q^2=M_Z^2$ the electroweak couplings $\hat e_q^2$ are such that
\be
\hat e_u^2\,<\,\hat e_d^2\,=\,\hat e_s^2, \qquad {\rm with}\qquad
\frac{\hat e_u^2}{\hat e_d^2}\approx \frac{3}{4}\cdot\label{ratio}
\ee
Hence there must exist a "magic" energy $Q_0^2$ where
\be
\hat e_u^2(Q_0^2)=\hat e_d^2(Q_0^2)=\hat e_s^2(Q_0^2).
\ee
At this energy
\be
\tilde\sigma_{e^+e^-}^{\pi^+} &\sim& D_u^{\pi^+} +D_d^{\pi^+}+D_s^{\pi^+}+
D_{\bar u}^{\pi^+} +D_{\bar d}^{\pi^+}+D_{\bar s}^{\pi^+}
\sim D_\Sigma^{\pi^+}
\ee
where $D_\Sigma^{\pi^+} = 2\left(  D_u^{\pi^+} +D_d^{\pi^+}+D_s^{\pi^+}\right)$
is the flavour singlet FF.

Unfortunately there is no data at the magic energy $\sqrt {Q^2_0}
= 78.4$ GeV, but it is close enough to $M_{Z^0}$ to be able to
argue that $D_{\Sigma}^{\pi^+}$ is very well determined by the
combination of FFs measured at the $Z^0$, namely,
 \be
D_{\Sigma}^{\pi^+} =\frac{43}{77} D_{measured}^{\pi^+ + \pi^-}\, (1.00\pm 0.02),
\ee
 i.e. $D_{\Sigma}^{\pi^+} $ is known accurately at $Q^2 =
M_Z^2$. However it has to be evolved down to $Q^2_{HERMES}$. This
involves mixing with the poorly known gluon FF $D_G^{\pi}$.
Consequently we end up knowing $D_{\Sigma}^{\pi^+} $ at
$Q^2_{HERMES}$ to $\pm 10\%$.

Knowing the value of  $D_{\Sigma}^{\pi^+} $ provides us with a
third equation so that we can now solve for the individual
$D_{q}^{\pi^+} $:
\be
&&D_u^{\pi^+}- D_d^{\pi^+}=
\frac{9\left(R_p^{\pi^+} -
R_p^{\pi^-}\right)\tilde{\sigma}_p^{DIS}}{4u_V -d_V}\nn\\
 &&D_u^{\pi^+}+ D_d^{\pi^+}=\frac{9\left(R_p^{\pi^+} +
 R_p^{\pi^-}\right)\tilde\sigma_p^{DIS} -2s\,D_{\Sigma}^{\pi^+}}{4(u+\bar u -s) +d+\bar d}
 \nn\\
 &&D_s^{\pi^+}=\frac{-18\left(R_p^{\pi^+} +
 R_p^{\pi^+}\right)\tilde\sigma_p^{DIS} +[4(u+\bar u) +d+\bar d]\,D_{\Sigma}^{\pi}}
 {2\,[4(u+\bar u -s) +d+\bar d\,]}
 \label{Ds}
 \ee
where $R^h \equiv \sigma^h/\sigma^{DIS} =
\tilde\sigma^h/\tilde\sigma^{DIS}$. Note that $D_u-D_d$ is
independent on $D_\Sigma$, that $D_u+D_d$ depends weakly on it,
and that $D_s$ is most sensitive to it. This is reflected in the
uncertainties shown on the $D_q^{\pi^+}$ derived from (\ref{Ds})
(see left of Fig. 2).
\figb

\section{Implications for the polarized parton densities}

At present the asymmetry data on $\pi^\pm$ SIDIS using a polarized
target are not yet available. In order, therefore, to study the
error propagation, we generate perfect  (error-free) fake proton
DIS and SIDIS asymmetry data, and then analyze it, following
closely the methods advocated by HERMES.

Thus we construct the flavour $f$ purities ($h=\pi^+,\pi^-$):
\be
 P_{q_f}^{h}(x)
 =\frac{e^2_{q_f}\,q_f(x)\,\int \,dz D_{q_f}^{h}(z)}
 {\sum_{f'}e^2_{q{f'}}\,q_{f'}(x)\,\int \,dz D_{q_{f'}}^{h}(z)},\qquad
P_{q_f}^{DIS}(x)
 =\frac{e^2_{q_f}\,q_f(x)}
 {\sum_{f'}e^2_{q_{f'}}\,q_{f'}(x)}\cdot
 \label{Purities}
 \ee
  The measured asymmetries integrated over $z$, are then given, in LO, by
\be
 <\Delta A_p^{h}(x)> \equiv \frac{\int dz
 \,\Delta\tilde\sigma^{h}(x,z)}{\int dz\,
 \tilde\sigma^{h}(x,z)}= \sum_{q,\bar q}\,P_{q}^{h}(x) \left(\frac{\Delta
 q(x)}{q(x)}\right),\qquad h=\pi^\pm, DIS\label{deltaAp}
 \ee

We have data on proton DIS and proton $\pi^\pm$ SIDIS at each $x$,
i.e. 3 pieces of information on the LHS of (\ref{deltaAp}). On the
RHS there are 6 unknown $\Delta q/q$. Again, following HERMES, we
make the somewhat bizarre assumption
\be
\Delta\bar u/\bar
u=\Delta\bar d/\bar d=\Delta s/
s=\Delta\bar s/\bar s \equiv\Delta
q_s/q_s\label{assumption}
\ee
and solve for $\Delta
u/u$, $\Delta d/d$ and $\Delta q_s/q_s$. The
results are shown in Fig. 2.
We see that even with our fake,
error-free DIS and $\pi^\pm $ SIDIS data, the uncertainties in the
FFs induce significant uncertainties in both  $\Delta d/d$
and $\Delta q_s/q_s$.

\section{Conclusions}

We have explained why the large uncertainty in our knowledge of
$\left(\Delta s(x) + \Delta\bar s(x)\right)$ and $\Delta G(x)$,
and the total lack of knowledge of the separate valence and sea
polarized densities, cannot be remedied by purely DIS experiments,
at least not until far in the future when a neutrino factory and
TESLA-N will be built.

All progress on the polarized PDs  thus rests
upon the HERMES and COMPASS polarized SIDIS  measurements, and the
utility of these, in turn rests upon an accurate knowledge of the
FFs. We have used a judicious combination of
HERMES $\pi^\pm$ multiplicity  and $e^+e^-$ data to evaluate
the 3 independent pion FFs $D_u^{\pi^+}$, $D_d^{\pi^+}$,
$D_s^{\pi^+}$ and their uncertainties. Using these we have
generated fake error-free DIS and SIDIS asymmetry data and
extracted the polarized PDs and their uncertainties
from these data, following the methods advocated by HERMES.


\section{Acknowledgements}
E.L. is grateful to the Leverhulme Trust for an Emeritus Fellowship, the
Foundation for Fundamental Research on Matter (FOM) and the Dutch
Organization for Scientific research (NWO), and to P. Mulders for
hospitality at the Vrije University, Amsterdam. E.C.'s work was
supported by the Bulgarian National Science Foundation, Grant
Ph-1010. S.K.'s work supported by the
National Science Foundation under Grant
PHY-0070443.



\end{document}